\documentstyle[11pt]{article}
\begin{document}
\baselineskip 18pt
\centerline{\Large\bf Y(so(5)) symmetry of the nonlinear}
\centerline{\Large\bf Schr$\ddot{o}$dinger model with four-components}

\vspace{3cm}
\centerline{ Hong-Biao Zhang$^{1,3}$, Mo-Lin Ge$^{1}$, Kang Xue$^{1,2}$}
\vspace{0.5cm}
{\small
\centerline{\bf 1. Theoretical Physics Division, Nankai Institute of Mathematics,}
\centerline{\bf Nankai University, Tianjin 300071, P.R.China}
\centerline{\bf 2. Physics Department, Northeast Normal University,}
\centerline{\bf Changchun,Jilin, 130024, P.R.China}
\centerline{\bf 3. Educational Institute of Jilin Province,}
\centerline{\bf Changchun,Jilin,  130022, P.R.China}
}
\vspace{2cm}

\centerline{\bf Abstract}

   The quantum nonlinear Schr$\ddot{o}$dinger(NLS) model with four-component
   fermions exhibits a $Y(so(5))$ symmetry when considered on an infintite
   interval. The constructed generators of Yangian are proved to satisfy
   the Drinfel'd formula and furthermore, the $RTT$ relation with the
   general form of rational R-matrix given by Yang-Baxterization associated
   with $so(5)$ algebraic structure.

\vspace{3cm}

\pagebreak

{\bf (I). Introduction}

  Many one-dimensional nonlinear models possess $Y(sl(2))$ symmetry. The
notable examples include Hubbard model$^{[1]}$ and its extension$^{[2]}$,
Haldance-shastry model$^{[3-8]}$ that are chain models and, as explored by
S.Murakami and M.Wadati, the NLS model with spins$^{[9]}$. In our knowledge
so far the shown Yangian symmtry exhibited by physical models is only for
$Y(sl(2))$. However, it has been proposed by S.C. Zhang et al. that the
antiferrromagnetic(AF) and superconducting(SC) phases of high-$T_c$ cuprates
are hopeful to be unified by an $so(5)$ symmetry principle$^{[10]}$. Possible
support for this proposal came from numerical investigations in models for
high-$T_c$ materials. In particular, it was shown that the low-energy
excitations can be classified in terms of an $so(5)$ symmetry multiplet
structure$^{[11,12]}$. Subsequently, extented Hubbard models and a two-leg
ladder model related to $so(5)$ symmetry have been introduced and analyzed
in detail$^{[13-15]}$. While these model don't exhibit a larger symmetry:
$Y(so(5))$ symmetry.
  In this paper we would like to show that the NLS model with four-component
fermions will have a larger symmtry: the Yangian symmetry $Y(so(5))$ which
is realized on a infinite interval. This work can be viewed as an extension
of Ref.[9] with mainly two different features: 1). the relations for $Y(so(5))$
are much more complicated than $Y(sl(2))$'s, and 2). in comparison to the
new realization of Yangian by Drinfel'd$^{[16]}$, we should complete the
calculations based on $RTT$ relation.
  This paper is organized as follows: In section (II). the commutativity
between the NLS model with $so(5)$ symmetry and the continuous realization
of $Y(so(5))$ is presented. In section (III). we recast the relations for
$Y(so(5))$ given by Drinfel'd to the simplest form with the help of Jacobi
identities. Then the presented operators $I^{(1)}_{ab}$ and $I^{(2)}_{ab}$
will be shown to satisfy $Y(so(5))$. In section (IV). we start from $RTT$
relation and show as the consequence to yield the simplified form of
$Y(so(5))$ that not only check our result, but also explicitly give the
"new realization" of $Y(so(5))$ given in Ref.[16].

{\bf (II). Hamiltonian and Realization  of $Y(so(5))$ }

In this section we will give explicit expression for the genrators of
$Y(so(5))$ commuting with the NLS model of four-component fermions.

The Hamiltonian for the NLS model of four-component fermions is given
by the following expression
\begin{equation}
\label{eq2.1}
H=\int \{\partial_x \psi_\alpha^+ \partial_x \psi_\alpha
+c{\psi_\beta^+} {\psi_\alpha^+} {\psi_\alpha} {\psi_\beta} \} dx
\end{equation}

Here $\psi_\alpha (x)(\alpha=1,2,3,4)$ denote fermionic field
operators which satisfy canonical anticommutation relations

$$
  [{\psi_\alpha^+ (x)},{\psi_\beta^+ (y)}]_+=0,\;\;\;
 [{\psi_\alpha (x),{\psi_\beta (y)}}]_+=0
$$
\begin{equation}
\label{eq2.2}
 [{\psi_\alpha (x)},\psi_\beta^+ (y)]_+=\delta _{\alpha \beta} \delta (x-y) \;\;\;\;\;\;\;\;\;\;\;
\end{equation}
We can construct Yangian generartors as follows (a,b=1,2,3,4,5)

$$
I_{ab}=I_{ab}^{(1)}=\int I_{ab}(x)dx  ,\;\;\;\;\;\;
I_{ab}(x)=\frac{1}{2}\psi_\alpha^+ (x)\Gamma_{\alpha \beta}^{ab}
\psi_\beta (x)
$$
\begin{equation}
\label{eq2.3}
J_{ab}=I_{ab}^{(2)}=-\frac{i}{2}\int dx \psi_\alpha^+ (x) \Gamma^{ab}_{\alpha \beta}
\partial_{x}\psi_\beta (x)-\frac{ic}{2} \int \int dxdy
\epsilon(x-y) I_{ac}(x)I_{cb}(y)
\end{equation}
where $\Gamma^{ab}=-i\Gamma^{a}\Gamma^{b}$ and $\Gamma^a$ are 4$\times$4 Dirac
matrices, $c=h$,in comparison to Ref.[10].
It can be checked that the set $\{I_{ab}^{(1)},I_{ab}^{(2)}\}$
commutes with the Hamiltonian of the NLS model: $[H,I_{ab}^{(1)}]
=[H,I_{ab}^{(2)}]=0$ or $[H,Y(so(5))]=0$, where the set $\{I_{ab},J_{ab}\}$
satisfies $Y(so(5))$. i.e., $Y(so(5))$ serves as a symmetry of the NLS model.
In momentum space, Eq.(\ref{eq2.3}) can be expressed as
$$
J_{ab}=I_{ab}^{(2)}=\int kI_{ab}(k,k)dk
-\frac{c}{\pi} \int \int dkdpdq
k^{-1} I_{ac}(k+p,p)I_{cb}(q,k+q)
$$
\begin{equation}
\label{eq2.4}
I_{ab}=I_{ab}^{(1)}=\int I_{ab}(k,k)dk  ,\;\;\;\;\;\;
I_{ab}(k,k')=\frac{1}{2}\psi_\alpha^+(k)\Gamma^{ab}_{\alpha \beta}\psi_\beta(k')
\end{equation}
For four-component fermions the field operator $\psi (k)=[c_{\sigma}(k),d^+_{\sigma}(-k)]^T$
($\sigma=\uparrow,\downarrow$ and $T$ means transport), then Eq.(\ref{eq2.1})
is changed into the following form
$$
H=\int \sum_{\sigma=\uparrow,\downarrow} \{(k^2+4c) c_\sigma^+(k)c_\sigma(k)
+d_\sigma^+(k)d_\sigma(k)-1\} dk
$$
$$
+\frac{c}{\pi}\int\int\int dkdk'dq
\{c_\uparrow^+(k+\frac{q}{2})c_\downarrow^+(-k+\frac{q}{2})
c_\downarrow (-k'+\frac{q}{2})c_\uparrow (k'+\frac{q}{2})
$$
$$
-c_\uparrow^+(k+\frac{q}{2})d_\uparrow^+(-k+\frac{q}{2})
d_\uparrow(-k'+\frac{q}{2})c_\uparrow (k'+\frac{q}{2})
$$
$$
-c_\uparrow^+(k+\frac{q}{2})d_\downarrow^+(-k+\frac{q}{2})
d_\downarrow(-k'+\frac{q}{2})c_\uparrow (k'+\frac{q}{2})
$$
$$
-c_\downarrow^+(k+\frac{q}{2})d_\uparrow^+(-k+\frac{q}{2})
d_\uparrow(-k'+\frac{q}{2})c_\downarrow (k'+\frac{q}{2})
$$
$$
-c_\downarrow^+(k+\frac{q}{2})d_\downarrow^+(-k+\frac{q}{2})
d_\downarrow^+(-k'+\frac{q}{2})c_\downarrow (k'+\frac{q}{2})
$$
\begin{equation}
\label{eq2.5}
+d_\uparrow(k+\frac{q}{2})d_\downarrow(-k+\frac{q}{2})
d_\downarrow^+(-k'+\frac{q}{2})d_\uparrow^+(k'+\frac{q}{2})\}
\end{equation}
Since $c_{\sigma}(k)$ and $d_{\sigma}(k)$ denote different fermions,
Eq.(\ref{eq2.5}) gives the pair interaction. Eq.(\ref{eq2.4}) can
be expressed in terms of Cartan-Weyl basis forms with

$$
E^{(1)}_{3}=\int E_{3}(k,k)dk,\;\;\;
E^{(1)}_{\pm}=\int E_{\pm}(k,k)dk,\;\;\;
F^{(1)}_{3}=\int F_{3}(k,k)dk,\;\;\;
$$
\begin{equation}
\label{eq2.6}
F^{(1)}_{\pm}=\int F_{\pm}(k,k)dk,\;\;\;
U^{(1)}_{\pm}=\int U_{\pm}(k,k)dk,\;\;\;
V^{(1)}_{\pm}=\int V_{\pm}(k,k)dk
\end{equation}
$$
E^{(2)}_3=\int kE_3(k,k)dk
-\frac{c}{\pi}\int \int \int dkdpdq k^{-1}
\{U_+(k+p,p)U_-(q,k+q) $$
$$
+V_+(k+p,p)V_-(q,k+q)+E_+(k+p,p)E_-(q,k+q)\}
$$
$$
E^{(2)}_{\pm}=\int kE_{\pm}(k,k)dk
\mp\frac{c}{\pi}\int \int \int dkdpdq k^{-1}
\{U_{\pm}(k+p,p)F_{\pm}(q,k+q)$$
$$
-V_{\pm}(k+p)F_{\mp}(q,k+q)+E_3(k+p,p)E_{\pm}(q,k+q)\}
$$
$$
F^{(2)}_3=\int kF_3(k,k)dk
-\frac{c}{\pi}\int \int \int dkdpdq k^{-1}
\{-U_+(k+p,p)U_-(q,k+q)
$$
$$
+V_+(k+p,p)V_-(q,k+q)+F_+(k+p,p)F_-(q,k+q)\}
$$
$$
F^{(2)}_{\pm}=\int kF_{\pm}(k,k)dk
\mp\frac{c}{\pi}\int \int \int dkdpdq k^{-1}
\{F_3(k+p,p)F_{\pm}(q,k+q)
$$
$$
+V_{\pm}(k+p,p)E_{\mp}(q,k+q)
-E_{\pm}(k+p,p)U_{\mp}(q,k+q)\}
$$
$$
U^{(2)}_{\pm}=\int kU_{\pm}(k,k)dk
\mp\frac{c}{\pi}\int \int \int dkdpdq k^{-1}
\{E_{\pm}(k+p,p)F_{\mp}(q,k+q)
$$
$$
+U_{\pm}(k+p,p)(F_3(q,k+q)-E_3(q,k+q)\}
$$
$$
V^{(2)}_{\pm}=\int kV_{\pm}(k,k)dk
\pm\frac{c}{\pi}\int \int \int dkdpdq k^{-1}
\{E_{\pm}(k+p,p)F_{\pm}(q,k+q)
$$
\begin{equation}
\label{eq2.7}
-(E_3(k+p,p)+F_3(k+p,p))V_{\pm}(q,k+q)\}
\end{equation}
where
$$
E_+(k,k')=\frac{1}{\sqrt{2}}(c_{\uparrow}^+(k)c_{\downarrow}(k')
-d_{\downarrow}(-k)d_{\uparrow}(-k')^+) \;\;\;\;\;\;\;\;\;\;\;\;\;\;\;\;\;\;\;\;\;\;\;\;\;\;\;\;\;\;
$$
$$
E_-(k,k')=\frac{1}{\sqrt{2}}(c_{\downarrow}^+(k)c_{\uparrow}(k')
-d_{\uparrow}(-k)d_{\downarrow}(-k')^+) \;\;\;\;\;\;\;\;\;\;\;\;\;\;\;\;\;\;\;\;\;\;\;\;\;\;\;\;\;\;\;
$$
$$
F_+(k,k')=\frac{-1}{\sqrt{2}}(d_{\uparrow}(-k)c_{\downarrow}(k')
+d_{\downarrow}(-k)c_{\uparrow}(k')) \;\;\;\;\;\;\;\;\;\;\;\;\;\;\;\;\;\;\;\;\;\;\;\;\;\;\;\;\;\;\;\;\;
$$
$$
F_-(k,k')=\frac{-1}{\sqrt{2}}(c_{\uparrow}^+(k)d_{\downarrow}^+(-k')
+c_{\downarrow}^+(k)d_{\uparrow}^+(-k')) \;\;\;\;\;\;\;\;\;\;\;\;\;\;\;\;\;\;\;\;\;\;\;\;\;\;\;\;\;\;\;\;
$$
$$
U_+(k,k')=-c_{\uparrow}^+(k)d_{\uparrow}^+(-k') ,\;\;\;\;\;\;\;\;\;\;\;\;\;
U_-(k,k')=-d_{\uparrow}(-k)c_{\uparrow}(k')
$$
$$
V_+(k,k')=-d_{\downarrow}(-k)c_{\downarrow}(k') ,\;\;\;\;\;\;\;\;\;\;\;\;\;\;
V_-(k,k')=-c_{\downarrow}^+(k)d_{\downarrow}^+(-k'))
$$
$$
E_3(k,k')=\frac{1}{2}(c_{\uparrow}^+(k)c_{\downarrow}(k')
-c_{\downarrow}^+(k)c_{\downarrow}(k')
-d_{\uparrow}(-k)d_{\uparrow}^+(-k')
+d_{\downarrow}(-k)d_{\downarrow}^+(-k') )
$$
$$
F_3(k,k')=\frac{1}{2}(-c_{\uparrow}^+(k)c_{\downarrow}(k')
-c_{\downarrow}^+(k)c_{\downarrow}(k')
+d_{\uparrow}(-k)d_{\uparrow}^+(-k')
+d_{\downarrow}(-k)d_{\downarrow}^+(-k') )
$$

The relations for $U^{(n)}_{\alpha}$, $V^{(n)}_{\alpha}$, $E^{(n)}_{\alpha}$,
$F^{(n)}_{\alpha}$ ($n=1,2$ and $\alpha=\pm,3$) and $I_{ab}^{(1)}$, $I_{ab}^{(2)}$
are given by
$$
E^{(n)}_3=\frac{1}{h^{n-1}}I^{(n)}_{23},\;\;
E^{(n)}_{\pm}=\frac{1}{\sqrt{2}h^{n-1}}(I^{(n)}_{34}\pm iI^{(n)}_{42}),\;\;
F^{(n)}_{\pm}=\frac{1}{\sqrt{2}h^{n-1}}(I^{(n)}_{45}\pm iI^{(n)}_{14})$$
$$
F^{(n)}_3=\frac{1}{h^{n-1}}I^{(n)}_{15},\;\;
U^{(n)}_{\pm}=\frac{1}{2h^{n-1}}\{(I^{(n)}_{31}\pm iI^{(n)}_{12})
-(I^{(n)}_{25}\pm iI^{(n)}_{35})\},\;\;\;\;\;\;\;\;  $$
\begin{equation}
\label{eq2.8}
V^{(n)}_{\pm}=\frac{1}{2h^{n-1}}\{(I^{(n)}_{31}\pm iI^{(n)}_{12})
+(I^{(n)}_{25}\pm iI^{(n)}_{35})\},\;\;\;\;\;\;\;\;\;\;\;\;\;\;\;\;\;\;\;
\end{equation}
Therefore, we have expressed the generators of $Y(so(5))$ in terms of
fermions including their pairs.
Equiping with the above relations we shall show that the $I_{ab}$ and
$J_{ab}$ given by Eq.(\ref{eq2.3}) really satisfy $Y(so(5))$.

{\bf (III). Simplication of the commutation relations of $Y(so(5))$ }

  The original commutation relations of Yangian $Y({\bf a})$ were given by
Drinfle'd$^{[17]}$ in the form:
\begin{equation}
\label{eq3.1}
[I_\lambda,I_\mu]=c_{\lambda \mu \nu}I_\nu, \;\;~~~~~~
[I_\lambda,J_\mu]=c_{\lambda \mu \nu}J_\nu,
\end{equation}
\begin{equation}
\label{eq3.2}
[J_\lambda,[J_\mu,I_\nu]]-[I_\lambda,[J_\mu,J_\nu]]=h^2a_{\lambda \mu \nu
\alpha \beta \gamma} \{I_\alpha,I_\beta,I_\gamma\},
\end{equation}
where eq.(\ref{eq3.2}) doesn't works only for $sl(2)$, and
\begin{equation}
\label{eq3.3}
[[J_\lambda,J_\mu],[I_\sigma,J_\tau]]+[[J_\sigma,J_\tau],[I_\lambda,J_\mu]]
 =h^2(a_{\lambda \mu \nu \alpha \beta \gamma}c_{\sigma \tau \nu}+a_{\sigma
 \tau \nu \alpha \beta \gamma} c_{\lambda \mu \nu}    )
\{I_\alpha,I_\beta,J_\gamma\},
\end{equation}
where $c_{\lambda \mu \nu}$ are structure constants of a simple Lie algebra
{\bf a}, $h$ is a constant and
$$a_{\lambda \mu \nu \alpha \beta \gamma}=\frac{1}{4!}
c_{\lambda \alpha \sigma}c_{\mu \beta \tau }
c_{\nu \gamma \rho}c_{\sigma \tau \rho}, $$
\begin{equation}
\label{eq3.4}
\{x_1,x_2,x_3\}=\sum_{i\ne j\ne k}x_ix_jx_k.
\end{equation}
Through the " new realization of Yangian" given in Ref.[16] by Drinfel'd the
isomorphic between $Y({\bf a})$ and $T^{(n)}_{ab}$ was established, where
$T_{ab}(u)$ are matrix elements of transfer matrix $T(u)$ obeying $RTT$
relation
\begin{equation}
\label{eq3.5}
\check{R}(u-v)(T(u) \otimes T(v))=(T(v) \otimes T(u))\check{R}(u-v)
\end{equation}
where $\check{R}(u)$ is the rational solution of YBE. The expansion of
$T(u)$ is made by:
\begin{equation}
\label{eq3.6}
T_{ab}(u)=\delta_{ab}+\sum_{n=1}^{\infty} u^{-n} T_{ab}^{(n)}.
\end{equation}
From the point of view of mathematics Ref.[16-18] together with the
representation theory of $Y(sl(2))$ by Chari and Pressley$^{[19]}$ have
solved the basic problems of Yangian associated with $sl(2)$ ($Y(sl(2))$).
For $sl(2)$ Eq.(\ref{eq3.3}) can be simplified to only one relation as was
shown in Ref.[8]. Now we can show that besides Eq.(\ref{eq3.1}) not all
of the relations in Eq.(\ref{eq3.2}) and Eq.(\ref{eq3.3}) are independent.
After tedious calculation we can prove that there is only one independent
relation:
\begin{equation}
\label{eq3.7}
[I_{23}^{(2)},I_{15}^{(2)}]=\frac{i}{4!}h^2 (\{I_{13}^{(1)},I_{42}^{(1)},I_{45}^{(1)}\}
+\{I_{12}^{(1)},I_{45}^{(1)},I_{34}^{(1)}\}
-\{I_{14}^{(1)},I_{42}^{(1)},I_{35}^{(1)}\}-\{I_{14}^{(1)},I_{34}^{(1)},I_{25}^{(1)}\})
\end{equation}
where $I_{23}^{(1)}$ and $I_{15}^{(1)}$ are the Cartan subset. All the other relations
in Eq.(\ref{eq3.2}) and Eq.(\ref{eq3.3}) can be generated on the basis of
Eq.(\ref{eq3.7}) by using Jocobi identities together with Eq.(\ref{eq3.1}).
Therfore, for $Y(so(5))$, Eq.(\ref{eq3.1})-Eq.(\ref{eq3.3}) can also be
expressed with Eq.(\ref{eq3.8})-Eq.(\ref{eq3.9})
\begin{equation}
\label{eq3.8}
[I_{ab}^{(1)},I_{cd}^{(1)}]=i(\delta _{bc}I_{ad}^{(1)} +\delta _{ad}I_{bc}^{(1)}
-\delta _{ac}I_{bd}^{(1)} -\delta _{bd}I_{ac}^{(1)})
\end{equation}
\begin{equation}
\label{eq3.9}
[I_{ab}^{(1)},I_{cd}^{(2)}]=i(\delta _{bc}I_{ad}^{(2)} +\delta _{ad}I_{bc}^{(2)}
-\delta _{ac}I_{bd}^{(2)} -\delta _{bd}I_{ac}^{(2)})
\end{equation}
$$
I_{ab}^{(1)}=-I_{ba}^{(1)}; \;\; I_{ab}^{(2)}=-I_{ba}^{(2)};\;\; (a,b,c,d=1,2,3,4,5)
$$
together with Eq.(\ref{eq3.7}). The proof is direct: substituting the structure
constants for $so(5)$ into Eq.(\ref{eq3.2}) and Eq.(\ref{eq3.3}), then carefully
using Jacobi identities together with Eq.(\ref{eq3.1}) and checking one by one
relation, finally based on Eq.(\ref{eq3.1}) we generate all the relations given
by Eq.(\ref{eq3.2}) and Eq.(\ref{eq3.3}) in terms of Eq.(\ref{eq3.7}) and Jacobi
identities. This process looks the extension for $Y(sl(2))$ as discussed in
Ref.[20]. The similar also occurs in $Y(su(3))$.
Direct check verifies that Eq.(\ref{eq2.3}) satisfies Eq.(\ref{eq3.7})-Eq.(\ref{eq3.9})
indeed.

{\bf (IV). RTT Relations and $Y(so(5))$}

We have proved that the operators $I_{ab}$ and $J_{ab}$ shown by
Eq.(\ref{eq2.3}) satisfies Eq.(\ref{eq3.7})-Eq.(\ref{eq3.9}), hence
Eq.(\ref{eq3.1})-Eq.(\ref{eq3.3}). To independently check the statement
and furthermore show the contensicy between Eq.(\ref{eq3.7}) and the
"new realization" of Yangian given by Drinfeld$^{[16]}$, we have to solve
$RTT$ relation for $so(5)$, i.e. starting from Eq.(\ref{eq3.5}) and
Eq.(\ref{eq3.6}) we shall give the general forms of $T_{ab}^{(n)}$ and
verify when $n=1,2$ they will give Eq.(\ref{eq3.7})-Eq.(\ref{eq3.9}).
For this goal we should begin with the rational R-matrix associated
with $so(5)$. For $so(N)$ the general form had been given in Ref.[21]
through Yang-Baxterization$^{[22]}$: $\check{R}(u)=f(u)\{u^2P+u(q_1P+
xA+q_2I)x+qI\}$, where $q_1=(1-\frac{N}{2})x, q_2=-x, q=q_1q_2$, $x$
is constant number and the elements of permutation matrix $P$, unit
matrix $I$ and blockdiag-matrix $A$ are $P^{ab}_{cd}=\delta_{ad} \delta_{bc},
I^{ab}_{cd}=\delta_{ac} \delta_{bd}$ and $A^{ab}_{cd}=\delta_{a,-b} \delta_{c,-d}$
when $a,b,c,d\in\{\frac{(-N+1)}{2},\frac{(-N+3)}{2},...,\frac{(N-3)}{2},\frac{(N-1)}{2}\}$.
For $so(5)$, $N=5$, $\check{R}$-matrix is changed into the following form
\begin{equation}
\label{eq4.1}
\check{R}(u)=f(u)\{u^2P+u(A-I-\frac{3}{2}P)x+\frac{3}{2}x^2I\}
\end{equation}
For our discussion we restrict ourselves to working in five-dimensional auxiliary
space, whereas each element $T_{ab}$ is a quantum operator, where $a,b=0,\pm1,\pm2$.
In comparison with Drinfeld$^{[16]}$, we express the transfer matrix $T(u)$ as
\begin{equation}
\label{eq4.2}
T_{ab}(u)= \delta_{ab}+\sum_{n=1}^{\infty} (\frac{x}{u})^n T_{ab}^{(n)}.
\end{equation}
Substituting Eq.(\ref{eq4.1}) and Eq.(\ref{eq4.2}) into Eq.(\ref{eq3.5}),
by direct calculation, we obtain the following relations

$$
[T^{(n+2)}_{bc},T^{(m)}_{ad}]-2[T^{(n+1)}_{bc},T^{(m+1)}_{ad}]
+[T^{(n)}_{bc},T^{(m+2)}_{ad}]+(T^{(n+1)}_{ic}T^{(m)}_{-id}
$$
$$
-T^{(n)}_{ic}T^{(m+1)}_{-id})\delta_{a,-b}
+(T^{(m+1)}_{ai}T^{(n)}_{b-i}-T^{(m)}_{ai}T^{(n+1)}_{b-i})\delta_{c,-d}
-\frac{3}{2}([T^{(n+1)}_{bc},T^{(m)}_{ad}]
$$
$$
-[T^{(n)}_{bc},T^{(m+1)}_{ad}])
-T^{(n+1)}_{ac}T^{(m)}_{bd}+T^{(n)}_{ac}T^{(m+1)}_{bd}
+T^{(m)}_{ac}T^{(n+1)}_{bd} $$
\begin{equation}
\label{eq4.3}
-T^{(m+1)}_{ac}T^{(n)}_{bd}+\frac{3}{2}(T^{(n)}_{ac}T^{(m)}_{bd}
-T^{(m)}_{ac}T^{(n)}_{bd})=0
\end{equation}
\begin{equation}
\label{eq4.4}
  [T^{(1)}_{bc},T^{(m)}_{ad}]+T^{(m)}_{-cd}\delta_{a,-b}
  -T^{(m)}_{a-b}\delta_{c,-d}-T^{(m)}_{bd}\delta_{ac}
  +T^{(m)}_{ac}\delta_{bd}=0
\end{equation}
\begin{equation}
\label{eq4.5}
  [T^{(n)}_{bc},T^{(1)}_{ad}]+T^{(n)}_{b-a}\delta_{c,-d}
  -T^{(n)}_{-dc}\delta_{a,-b}-T^{(n)}_{bd}\delta_{ac}
  +T^{(n)}_{ac}\delta_{bd}=0
\end{equation}
From Eq.(\ref{eq4.4}) its exchange ($a\leftrightarrow b , c\leftrightarrow
d , n\leftrightarrow m $) leads to Eq.(\ref{eq4.5}). It indicates that
Eq.(\ref{eq4.4}) is exactly the same as Eq.(\ref{eq4.5}). For the convenience
of the following calculation, we define $T_{ab}^{(n)}$ as follows
$$
T^{(n)}_{22}-T^{(n)}_{-2-2}=2E^{(n)}_3 ,\;\;\;
T^{(n)}_{22}+T^{(n)}_{-2-2}=2\tilde{E}^{(n)}_3 ,\;\;
T^{(n)}_{21}-T^{(n)}_{-1-2}=2U^{(n)}_+
$$
$$
T^{(n)}_{21}+T^{(n)}_{-1-2}=2\tilde{U}^{(n)}_+ ,\;\;
T^{(n)}_{20}-T^{(n)}_{0-2}=2E^{(n)}_+ ,\;\;\;
T^{(n)}_{20}+T^{(n)}_{0-2}=2\tilde{E}^{(n)}_+
$$
$$
T^{(n)}_{2-1}-T^{(n)}_{1-2}=2V^{(n)}_+ ,\;\;\;
T^{(n)}_{2-1}+T^{(n)}_{1-2}=2\tilde{V}^{(n)}_+ ,\;\;
T^{(n)}_{12}-T^{(n)}_{-2-1}=2U^{(n)}_-
$$
$$
T^{(n)}_{12}+T^{(n)}_{-2-1}=2\tilde{U}^{(n)}_- ,\;\;\;
T^{(n)}_{11}-T^{(n)}_{-1-1}=2F^{(n)}_3 ,\;\;\;
T^{(n)}_{11}+T^{(n)}_{-1-1}=2\tilde{F}^{(n)}_3
$$
$$
T^{(n)}_{10}-T^{(n)}_{0-1}=2F^{(n)}_+ ,\;\;\;
T^{(n)}_{10}+T^{(n)}_{0-1}=2\tilde{F}^{(n)}_+ ,\;\;\;
T^{(n)}_{02}-T^{(n)}_{-20}=2E^{(n)}_-
$$
$$
T^{(n)}_{02}+T^{(n)}_{-20}=2\tilde{E}^{(n)}_- ,\;\;\;
T^{(n)}_{01}-T^{(n)}_{-10}=2F^{(n)}_- ,\;\;\;
T^{(n)}_{01}+T^{(n)}_{-10}=2\tilde{F}^{(n)}_-
$$
$$
T^{(n)}_{1-1}=Y^{(n)}_+ ,\;\;\;\;\;\;\;
T^{(n)}_{-11}=Y^{(n)}_- ,\;\;\;\;\;\;\;
T^{(n)}_{-22}=X^{(n)}_- ,\;\;\;\;\;\;\;
T^{(n)}_{2-2}=X^{(n)}_+ \;\;\;
$$
\begin{equation}
\label{eq4.6}
T^{(n)}_{-12}-T^{(n)}_{-21}=2V^{(n)}_- ,\;\;\;
T^{(n)}_{-12}+T^{(n)}_{-21}=2\tilde{V}^{(n)}_- ,\;\;\;
T^{(n)}_{00}=I^{(n)}_{0}
\end{equation}
This defination of $T_{ab}^{(n)}$ is expressed in terms of matrix
form with
\begin{equation}
\label{eq4.7}
T_{ab}^{(n)}=\left(
\begin{array}{ccccc}
\tilde{E}^{(n)}_3+E^{(n)}_3&
\tilde{U}^{(n)}_++U^{(n)}_+&
\tilde{E}^{(n)}_++E^{(n)}_+&
\tilde{V}^{(n)}_++V^{(n)}_+&
X^{(n)}_+\\
\tilde{U}^{(n)}_-+U^{(n)}_-&
\tilde{F}^{(n)}_3+F^{(n)}_3&
\tilde{F}^{(n)}_++F^{(n)}_+&
Y^{(n)}_+&
\tilde{V}^{(n)}_+-V^{(n)}_+\\
\tilde{E}^{(n)}_-+E^{(n)}_-&
\tilde{F}^{(n)}_-+F^{(n)}_-&
I^{(n)}_0&
\tilde{F}^{(n)}_+-F^{(n)}_+&
\tilde{E}^{(n)}_+-E^{(n)}_+\\
\tilde{V}^{(n)}_-+V^{(n)}_-&
Y^{(n)}_-&
\tilde{F}^{(n)}_--F^{(n)}_-&
\tilde{F}^{(n)}_3-F^{(n)}_3&
\tilde{U}^{(n)}_+-U^{(n)}_+\\
X^{(n)}_-&
\tilde{V}^{(n)}_--V^{(n)}_-&
\tilde{E}^{(n)}_--E^{(n)}_-&
\tilde{U}^{(n)}_--U^{(n)}_-&
\tilde{E}^{(n)}_3-E^{(n)}_3\\
\end{array}   \right)
\end{equation}
For Eq.(\ref{eq4.5}), when $n=1$, substituting Eqs.(\ref{eq4.6})
into Eq.(\ref{eq4.5}), by tremendous calculation, we can obtain
the algebraic relations in Appendix Eqs.(30) and the
following constraint conditions
\begin{equation}
\label{eq4.8}
X^{(1)}_{\pm}=Y^{(1)}{\pm}
=\tilde{U}^{(1)}_{\pm}
=\tilde{V}^{(1)}_{\pm}
=\tilde{E}^{(1)}_{\pm}
=\tilde{F}^{(1)}_{\pm}=0,\;\;
\tilde{E}^{(1)}_3
=\tilde{F}^{(1)}_3
=I^{(1)}_0
\end{equation}
Therefore, $T^{(1)}$ is expressed in terms of the generators
of Lie algebra $so(5)$ with
\begin{equation}
\label{eq4.9}
T_{ab}^{(1)}=\left(
\begin{array}{ccccc}
I^{(1)}_0+E^{(1)}_3&
U^{(1)}_+&
E^{(1)}_+&
V^{(1)}_+&
0\\
U^{(1)}_-&
I^{(1)}_0+F^{(1)}_3&
F^{(1)}_+&
0&
-V^{(1)}_+\\
E^{(1)}_-&
F^{(1)}_-&
I^{(n)}_0&
-F^{(1)}_+&
-E^{(1)}_+\\
V^{(1)}_-&
0&
-F^{(1)}_-&
I^{(1)}_0-F^{(1)}_3&
-U^{(1)}_+\\
0&
-V^{(1)}_-&
-E^{(1)}_-&
-U^{(1)}_-&
I^{(1)}_0-E^{(1)}_3\\
\end{array}   \right)
\end{equation}
where $I^{(1)}_0$ is a Casimor operator. While for arbitray $n$,
substituting Eqs.(\ref{eq4.6}) into Eq.(\ref{eq4.5}) we can calculate
commutators between $T^{(1)}$ and $T^{(n)}$ in Appendix Eqs.(33) and
Eqs.(34).

When taking $n=1$, substituting Eqs.(\ref{eq4.6}) into Eq.(\ref{eq4.3})
by direct calculations, we can obtain the following constraint conditions
$$
\tilde{E}^{(2)}_3-I^{(2)}_0
=\frac{1}{4}([U^{(1)}_+,U^{(1)}_-]_+
+[V^{(1)}_+,V^{(1)}_-]_+
-2[F^{(1)}_+,F^{(1)}_-]_+
-[E^{(1)}_+,E^{(1)}_-]_+ + 2{E^{(1)}_3}^2)
$$
$$
\tilde{F}^{(2)}_3-I^{(2)}_0
=\frac{1}{4}([U^{(1)}_+,U^{(1)}_-]_+
+[V^{(1)}_+,V^{(1)}_-]_+
-2[E^{(1)}_+,E^{(1)}_-]_+
-[F^{(1)}_+,F^{(1)}_-]_+ + 2{F^{(1)}_3}^2)
$$
$$
\tilde{U}^{(2)}_{\pm}=\frac{1}{4}([E^{(1)}_{\pm},F^{(1)}_{\mp}]_+
+[E^{(1)}_3,U^{(1)}_{\pm}]_+ +[F^{(1)}_3,U^{(1)}_{\pm}]_+)
$$
$$
\tilde{V}^{(2)}_{\pm}=\frac{1}{4}(-[E^{(1)}_{\pm},F^{(1)}_{\pm}]_+
+[E^{(1)}_3,V^{(1)}_{\pm}]_+ -[F^{(1)}_3,V^{(1)}_{\pm}]_+)
$$
$$
\tilde{E}^{(2)}_{\pm}=\frac{1}{4}([E^{(1)}_3,E^{(1)}_{\pm}]_+
-[V^{(1)}_{\pm},F^{(1)}_{\mp}]_+ +[F^{(1)}_{\pm},U^{(1)}_{\pm}]_+)
$$
$$
\tilde{F}^{(2)}_{\pm}=\frac{1}{4}([E^{(1)}_{\mp},V^{(1)}_{\pm}]_+
+[E^{(1)}_{\pm},U^{(1)}_{\mp}]_+ +[F^{(1)}_3,F^{(1)}_{\pm}]_+)
$$
$$
X^{(2)}_{\pm}=-\frac{1}{4}([E^{(1)}_{\pm},E^{(1)}_{\pm}]_+
+2[U^{(1)}_{\pm},V^{(1)}_{\pm}]_+ )
$$
\begin{equation}
\label{eq4.10}
Y^{(2)}_{\pm}=-\frac{1}{4}([F^{(1)}_{\pm},F^{(1)}_{\pm}]_+
-2[U^{(1)}_{\mp},V^{(1)}_{\pm}]_+ )
\end{equation}

Substituiting Eqs.(\ref{eq4.6}) into Eq.(\ref{eq4.3}), by using the
constraint conditions Eqs.(\ref{eq4.10}), Eqs.(\ref{eq4.8}) can be
reduced the following independent relations

$$
[E^{(n)}_3,F^{(2)}_3]+\frac{1}{4}([U^{(1)}_+,\tilde{U}^{(n)}_-]_+
-[U^{(1)}_-,\tilde{U}^{(n)}_+]_+ +[V^{(1)}_-,\tilde{V}^{(n)}_+]_+
$$
$$
-[V^{(1)}_+,\tilde{V}^{(n)}_-]_+)=0
$$
$$
[E^{(n)}_3,F^{(2)}_3]=[E^{(2)}_3,F^{(n)}_3]
$$
$$
[E^{(n)}_3,E^{(2)}_3]+\frac{1}{4}([U^{(1)}_-,\tilde{U}^{(n)}_+]_+
-[U^{(1)}_+,\tilde{U}^{(n)}_-]_+ +[V^{(1)}_-,\tilde{V}^{(n)}_+]_+
$$
$$
-[V^{(1)}_+,\tilde{V}^{(n)}_-]_+ +[E^{(1)}_-,\tilde{E}^{(n)}_+]_+
-[E^{(1)}_+,\tilde{E}^{(n)}_-]_+)=0
$$
$$
[F^{(n)}_3,F^{(2)}_3]+\frac{1}{4}([U^{(1)}_+,\tilde{U}^{(n)}_-]_+
-[U^{(1)}_-,\tilde{U}^{(n)}_+]_+ +[V^{(1)}_+,\tilde{V}^{(n)}_-]_+
$$
$$
-[V^{(1)}_-,\tilde{V}^{(n)}_+]_+ +[F^{(1)}_-,\tilde{F}^{(n)}_+]_+
-[F^{(1)}_+,\tilde{F}^{(n)}_-]_+)=0
$$
$$
[I^{(n)}_0,I^{(2)}_0]+([E^{(1)}_-,\tilde{E}^{(n)}_+]_+
-[E^{(1)}_+,\tilde{E}^{(n)}_-]_+ +[F^{(1)}_-,\tilde{F}^{(n)}_+]_+
$$
$$
-[F^{(1)}_+,\tilde{F}^{(n)}_-]_+)=0
$$
$$
[E^{(n)}_3,I^{(2)}_0]+\frac{1}{2}([E^{(1)}_+,E^{(n)}_-]_+
-[E^{(n)}_+,E^{(1)}_-]_+)=0
$$
$$
[F^{(n)}_3,I^{(2)}_0]+\frac{1}{2}([F^{(1)}_+,F^{(n)}_-]_+
-[F^{(1)}_-,F^{(n)}_+]_+)=0
$$
$$
[\tilde{E}^{(n)}_3,F^{(2)}_3]+\frac{1}{4}([U^{(1)}_+,U^{(n)}_-]_+
-[U^{(1)}_-,U^{(n)}_+]_+ +[V^{(1)}_-,V^{(n)}_+]_+
$$
$$
-[V^{(1)}_+,V^{(n)}_-]_+)=0
$$
$$
[\tilde{F}^{(n)}_3,E^{(2)}_3]+\frac{1}{4}([U^{(n)}_+,U^{(1)}_-]_+
-[U^{(n)}_-,U^{(1)}_+]_+ +[V^{(1)}_-,V^{(n)}_+]_+
$$
$$
-[V^{(1)}_+,V^{(n)}_-]_+)=0
$$
$$
[\tilde{E}^{(n)}_3,E^{(2)}_3]+\frac{1}{4}([U^{(1)}_-,U^{(n)}_+]_+
-[U^{(1)}_+,U^{(n)}_-]_+ +[V^{(1)}_-,V^{(n)}_+]_+
$$
$$
-[V^{(1)}_+,V^{(n)}_-]_+ +[E^{(1)}_-,E^{(n)}_+]_+
-[E^{(1)}_+,E^{(n)}_-]_+)=0
$$
$$
[\tilde{F}^{(n)}_3,F^{(2)}_3]+\frac{1}{4}([U^{(1)}_+,U^{(n)}_-]_+
-[U^{(1)}_-,U^{(n)}_+]_+ +[V^{(1)}_-,V^{(n)}_+]_+
$$
\begin{equation}
\label{eq4.11}
-[V^{(1)}_+,V^{(n)}_-]_+ +[F^{(1)}_-,F^{(n)}_+]_+
-[F^{(1)}_+,F^{(n)}_-]_+)=0
\end{equation}
$$
E^{(n+1)}_3=[E^{(n)}_+,E^{(2)}_-]+\frac{1}{4}([I^{(n)}_0,E^{(1)}_3]_+
-2[I^{(1)}_0,E^{(n)}_3]_+
$$
$$-[E^{(1)}_-,\tilde{E}^{(n)}_+]_+
-[F^{(1)}_-,\tilde{F}^{(n)}_+]_+ +[F^{(1)}_+,\tilde{F}^{(n)}_-]_+
+[U^{(1)}_-,\tilde{U}^{(n)}_+]_+
$$
$$
+[V^{(1)}_-,\tilde{V}^{(n)}_+]_+
+[E^{(1)}_3,\tilde{E}^{(n)}_3]_+)
$$
$$
F^{(n+1)}_3=[F^{(n)}_+,F^{(2)}_-]+\frac{1}{4}([I^{(n)}_0,F^{(1)}_3]_+
-2[I^{(1)}_0,F^{(n)}_3]_+ -[F^{(1)}_-,\tilde{F}^{(n)}_+]_+
$$
$$
-[E^{(1)}_-,\tilde{E}^{(n)}_+]_+ +[E^{(1)}_+,\tilde{E}^{(n)}_-]_+
+[U^{(1)}_+,\tilde{U}^{(n)}_-]_+ -[V^{(1)}_-,\tilde{V}^{(n)}_+]_+
$$
$$
+[F^{(1)}_3,\tilde{F}^{(n)}_3]_+)
$$
$$
\tilde{E}^{(n+1)}_3-I^{(n+1)}_0=[\tilde{E}^{(n)}_+,E^{(2)}_-]
+\frac{1}{4}(2[I^{(n)}_0,I^{(1)}_0]_+-2[I^{(1)}_0,\tilde{E}^{(n)}_3]_+
$$
$$
-[E^{(1)}_-,E^{(n)}_+]_+ -[F^{(1)}_-,F^{(n)}_+]_+ -[F^{(1)}_+,F^{(n)}_-]_+
+[U^{(1)}_-,U^{(n)}_+]_+
$$
$$+[V^{(1)}_-,V^{(n)}_+]_+
+[E^{(1)}_3,E^{(n)}_3]_+)
$$
$$
\tilde{F}^{(n+1)}_3-I^{(n+1)}_0=[\tilde{F}^{(n)}_+,F^{(2)}_-]
+\frac{1}{4}(2[I^{(n)}_0,I^{(1)}_0]_+ -2[I^{(1)}_0,\tilde{F}^{(n)}_3]_+
$$
$$
-[F^{(1)}_-,F^{(n)}_+]_+ -[E^{(1)}_-,E^{(n)}_+]_+ -[E^{(1)}_+,E^{(n)}_-]_+
+[U^{(1)}_+,U^{(n)}_-]_+
$$
\begin{equation}
\label{eq4.12}
+[V^{(1)}_-,V^{(n)}_+]_+
+[F^{(1)}_3,F^{(n)}_3]_+)
\end{equation}
where Eqs.(\ref{eq4.11}) are the constraint conditions, Eqs.(\ref{eq4.12})
are the iterative relation. All the other relation in Eq.(\ref{eq4.3})
can be generated on the basis of Eqs.(\ref{eq4.11}) and Eqs.(\ref{eq4.12})
by making use of Jocobi indentities together with Eqs.(30), Eqs.(31) and
constraint condition Eqs.(\ref{eq4.11}). Taking $n=2$ in Eqs.(\ref{eq4.11})
and Eqs.(\ref{eq4.12}), we derive only one independent relation which
corresponds to Eq.(\ref{eq3.7}) in Appendix Eq.(32). It is emphasized that
because of the iterative relation Eqs.(\ref{eq4.12}), only $T^{(1)}$ and
$T^{(2)}$ are basic ones.
To satisfy all the relations with $T^{(n)} (n\geq3)$ it is enough to look
for the constraints to $T^{(3)}$ that in turn to provide the constraints
to $T^{(2)}$ itself. In short summary, we have verified the equivalence between
Eq.(\ref{eq3.7})-Eq(\ref{eq3.9}) and the expansions of $RTT$ relations for
$m=0, n\leq2$ for $Y(so(5))$.

{\bf (V). Conclusion}

In this paper we have simplified the relations satisfied by $Y(so(5))$
to Eq.(\ref{eq3.7})-Eq.(\ref{eq3.9}) and shown their correspindence in
$RTT$ relation.
Also we have made the  realization of $Y(so(5))$ in the NLS model of
four-component fermions and shown Yangian symmetry for such a NLS model.
This work is in part supported by NSF of China.

\centerline{\bf APPENDIX}

In Cartan-Weyl basis Eq.(\ref{eq2.8}), Eq.(\ref{eq3.8})-Eq.(\ref{eq3.9})
are changed in the following forms
$$
[E^{(1)}_3,U^{(1)}_{\pm}]=\pm U^{(1)}_{\pm} ,\;\;
[E^{(1)}_3,E^{(1)}_{\pm}]=\pm E^{(1)}_{\pm} ,\;\;
[E^{(1)}_3,V^{(1)}_{\pm}]=\pm V^{(1)}_{\pm} ,\;\;
[E^{(1)}_3,F^{(1)}_{\alpha}]=0 ,\;\;\;\;\;\;
$$
$$
[E^{(1)}_{\pm},U^{(1)}_{\mp}]=\mp F^{(1)}_{\pm},\;\;
[E^{(1)}_{\pm},U^{(1)}_{\pm}]=0 ,\;\;\;\;
[U^{(1)}_{\pm},V^{(1)}_{\pm}]=0 ,\;\;\;\;
[V^{(1)}_+,V^{(1)}_-]=E^{(1)}_3+F^{(1)}_3 ,\;
$$
$$
[U^{(1)}_{\pm},V^{(1)}_{\mp}]=0 ,\;\;\;\;
[E^{(1)}_{\pm},V^{(1)}_{\pm}]=0 ,\;\;\;\;\;
[E^{(1)}_{\pm},V^{(1)}_{\mp}]=\pm F^{(1)}_{\mp} ,\;\;
[U^{(1)}_+,U^{(1)}_-]=E^{(1)}_3-F^{(1)}_3
$$
$$
[E^{(1)}_{\pm},F^{(1)}_{\pm}]=\mp V^{(1)}_{\pm} ,\;\;
[E^{(1)}_{\pm},F^{(1)}_{\mp}]=\pm U^{(1)}_{\pm} ,\;\;
[F^{(1)}_{\pm},V^{(1)}_{\mp}]=\mp E^{(1)}_{\mp} ,\;\;
[F^{(1)}_{\pm},V^{(1)}_{\pm}]=0 ,\;\;\;\;
$$
$$
[F^{(1)}_{\pm},U^{(1)}_{\pm}]=\mp E^{(1)}_{\pm} ,\;\;
[F^{(1)}_{\pm},U^{(1)}_{\mp}]=0 ,\;\;\;\;
[F^{(1)}_3,U^{(1)}_{\pm}]=\mp U^{(1)}_{\pm} ,\;\;
[F^{(1)}_3,V^{(1)}_{\pm}]=\pm V^{(1)}_{\pm}
$$
\begin{equation}
\label{eq30}
[F^{(1)}_3,F^{(1)}_{\pm}]=\pm F^{(1)}_{\pm} ,\;\;
[F^{(1)}_3,E^{(1)}_{\pm}]=0 ,\;\;\;\;
[E^{(1)}_+,E^{(1)}_-]=E^{(1)}_3 ,\;\;
[F^{(1)}_+,F^{(1)}_-]=F^{(1)}_3
\end{equation}

$$
[E^{(2)}_3,U^{(1)}_{\pm}]=\pm U^{(2)}_{\pm}=[E^{(1)}_3,U^{(2)}_{\pm}] ,\;\;
[E^{(2)}_3,E^{(1)}_{\pm}]=\pm E^{(2)}_{\pm}=[E^{(1)}_3,E^{(2)}_{\pm}]
$$
$$
[E^{(2)}_3,V^{(1)}_{\pm}]=\pm V^{(2)}_{\pm}=[E^{(1)}_3,V^{(2)}_{\pm}] ,\;\;\;\;\;
[E^{(2)}_3,F^{(1)}_{\alpha}]=0=[E^{(1)}_3,F^{(2)}_{\alpha}]
$$
$$
[E^{(2)}_{\pm},U^{(1)}_{\mp}]=\mp F^{(2)}_{\pm}=[E^{(1)}_{\pm},U^{(2)}_{\mp}] ,\;\;\;\;\;
[E^{(2)}_{\pm},U^{(1)}_{\pm}]=0=[E^{(1)}_{\pm},U^{(2)}_{\pm}]
$$
$$
[U^{(2)}_{\pm},V^{(1)}_{\pm}]=0=[U^{(1)}_{\pm},V^{(2)}_{\pm}] ,\;\;\;\;\;\;\;\;
[U^{(2)}_{\pm},V^{(1)}_{\mp}]=0=[U^{(1)}_{\pm},V^{(2)}_{\mp}]
$$
$$
[E^{(2)}_{\pm},V^{(1)}_{\pm}]=0=[E^{(1)}_{\pm},V^{(2)}_{\pm}] ,\;\;\;\;\;
[E^{(2)}_{\pm},V^{(1)}_{\mp}]=\pm F^{(2)}_{\mp}=[E^{(1)}_{\pm},V^{(2)}_{\mp}]
$$
$$
[U^{(2)}_{\pm},U^{(1)}_{\pm}]=0=[V^{(2)}_{\pm},V^{(1)}_{\pm}] ,\;\;\;\;
[V^{(2)}_+,V^{(1)}_-]=E^{(n)}_3+F^{(2)}_3=[V^{(1)}_+,V^{(2)}_-]
$$
$$
[E^{(2)}_{\alpha},E^{(1)}_{\alpha}]=0=[F^{(2)}_{\alpha},F^{(1)}_{\alpha}] ,\;\;\;\;
[U^{(2)}_+,U^{(1)}_-]=E^{(n)}_3-F^{(2)}_3=[U^{(1)}_+,U^{(2)}_-]
$$
$$
[E^{(2)}_{\pm},F^{(1)}_{\pm}]=\mp V^{(2)}_{\pm}=[E^{(1)}_{\pm},F^{(2)}_{\pm}] ,\;\;
[E^{(2)}_{\pm},F^{(1)}_{\mp}]=\pm U^{(2)}_{\pm}=[E^{(1)}_{\pm},F^{(2)}_{\mp}]
$$
$$
[F^{(2)}_{\pm},V^{(1)}_{\mp}]=\mp E^{(2)}_{\mp}=[F^{(1)}_{\pm},V^{(2)}_{\mp}] ,\;\;\;\;\;
[F^{(2)}_{\pm},V^{(1)}_{\pm}]=0=[F^{(1)}_{\pm},V^{(2)}_{\pm}]
$$
$$
[F^{(2)}_{\pm},U^{(1)}_{\pm}]=\mp E^{(2)}_{\pm}=[F^{(1)}_{\pm},U^{(2)}_{\pm}] ,\;\;\;\;\;
[F^{(2)}_{\pm},U^{(1)}_{\mp}]=0=[F^{(1)}_{\pm},U^{(2)}_{\mp}]
$$
$$
[F^{(2)}_3,U^{(1)}_{\pm}]=\mp U^{(2)}_{\pm}=[F^{(1)}_3,U^{(2)}_{\pm}] ,\;\;
[F^{(2)}_3,V^{(1)}_{\pm}]=\pm V^{(2)}_{\pm}=[F^{(1)}_3,V^{(2)}_{\pm}]
$$
$$
[F^{(2)}_3,F^{(1)}_{\pm}]=\pm F^{(2)}_{\pm}=[F^{(1)}_3,F^{(2)}_{\pm}] ,\;\;\;\;\;
[F^{(2)}_3,E^{(1)}_{\pm}]=0=[F^{(1)}_3,E^{(2)}_{\pm}]
$$
\begin{equation}
\label{eq31}
[E^{(2)}_+,E^{(1)}_-]=E^{(2)}_3=[E^{(1)}_+,E^{(2)}_-] ,\;\;
[F^{(2)}_+,F^{(1)}_-]=F^{(2)}_3=[F^{(1)}_+,E^{(2)}_-]
\end{equation}
But Eq.(\ref{eq3.7}) is changed into as follows

$$
[E^{(2)}_3,F^{(2)}_3]=\frac{1}{4}(U^{(1)}_-E^{(1)}_+F^{(1)}_-
-F^{(1)}_+E^{(1)}_-U^{(1)}_+ +V^{(1)}_-E^{(1)}_+F^{(1)}_+
$$
\begin{equation}
\label{32}
-F^{(1)}_-E^{(1)}_- V^{(1)}_+)  \;\;\;\;\;\;\;\;\;\;\;\;\;\;\;
\end{equation}
All the other relation other than Eq.(\ref{eq3.2})-Eq.(\ref{eq3.3})
can also be generated on the basis of Eq.(32) by using Jocobi identities
together with Eqs.(30) and Eqs.(31). Namely, this is the defination of $Y(so(5))$.

Substituting Eqs.(\ref{eq4.6}) into Eq.(\ref{eq4.5}), by tremendous calculation,
Eq.(\ref{eq4.5}) is changed into the following algebraic relations:
$$
[E^{(n)}_3,U^{(1)}_{\pm}]=\pm U^{(n)}_{\pm}=[E^{(1)}_3,U^{(n)}_{\pm}] ,\;\;
[E^{(n)}_3,E^{(1)}_{\pm}]=\pm E^{(n)}_{\pm}=[E^{(1)}_3,E^{(n)}_{\pm}]
$$
$$
[E^{(n)}_3,V^{(1)}_{\pm}]=\pm V^{(n)}_{\pm}=[E^{(1)}_3,V^{(n)}_{\pm}] ,\;\;\;\;\;
[E^{(n)}_3,F^{(1)}_{\alpha}]=0=[E^{(1)}_3,F^{(n)}_{\alpha}]
$$
$$
[E^{(n)}_{\pm},U^{(1)}_{\mp}]=\mp F^{(n)}_{\pm}=[E^{(1)}_{\pm},U^{(n)}_{\mp}] ,\;\;\;\;\;
[E^{(n)}_{\pm},U^{(1)}_{\pm}]=0=[E^{(1)}_{\pm},U^{(n)}_{\pm}]
$$
$$
[U^{(n)}_{\pm},V^{(1)}_{\pm}]=0=[U^{(1)}_{\pm},V^{(n)}_{\pm}] ,\;\;\;\;\;\;\;\;
[U^{(n)}_{\pm},V^{(1)}_{\mp}]=0=[U^{(1)}_{\pm},V^{(n)}_{\mp}]
$$
$$
[E^{(n)}_{\pm},V^{(1)}_{\pm}]=0=[E^{(1)}_{\pm},V^{(n)}_{\pm}] ,\;\;\;\;\;
[E^{(n)}_{\pm},V^{(1)}_{\mp}]=\pm F^{(n)}_{\mp}=[E^{(1)}_{\pm},V^{(n)}_{\mp}]
$$
$$
[U^{(n)}_{\pm},U^{(1)}_{\pm}]=0=[V^{(n)}_{\pm},V^{(1)}_{\pm}] ,\;\;\;\;
[V^{(n)}_+,V^{(1)}_-]=E^{(n)}_3+F^{(n)}_3=[V^{(1)}_+,V^{(n)}_-]
$$
$$
[E^{(n)}_{\alpha},E^{(1)}_{\alpha}]=0=[F^{(n)}_{\alpha},F^{(1)}_{\alpha}] ,\;\;\;\;\;
[U^{(n)}_+,U^{(1)}_-]=E^{(n)}_3-F^{(n)}_3=[U^{(1)}_+,U^{(n)}_-]
$$
$$
[E^{(n)}_{\pm},F^{(1)}_{\pm}]=\mp V^{(n)}_{\pm}=[E^{(1)}_{\pm},F^{(n)}_{\pm}] ,\;\;
[E^{(n)}_{\pm},F^{(1)}_{\mp}]=\pm U^{(n)}_{\pm}=[E^{(1)}_{\pm},F^{(n)}_{\mp}]
$$
$$
[F^{(n)}_{\pm},V^{(1)}_{\mp}]=\mp E^{(n)}_{\mp}=[F^{(1)}_{\pm},V^{(n)}_{\mp}] ,\;\;\;\;\;
[F^{(n)}_{\pm},V^{(1)}_{\pm}]=0=[F^{(1)}_{\pm},V^{(n)}_{\pm}]
$$
$$
[F^{(n)}_{\pm},U^{(1)}_{\pm}]=\mp E^{(n)}_{\pm}=[F^{(1)}_{\pm},U^{(n)}_{\pm}] ,\;\;\;\;\;
[F^{(n)}_{\pm},U^{(1)}_{\mp}]=0=[F^{(1)}_{\pm},U^{(n)}_{\mp}]
$$
$$
[F^{(n)}_3,U^{(1)}_{\pm}]=\mp U^{(n)}_{\pm}=[F^{(1)}_3,U^{(n)}_{\pm}] ,\;\;
[F^{(n)}_3,V^{(1)}_{\pm}]=\pm V^{(n)}_{\pm}=[F^{(1)}_3,V^{(n)}_{\pm}]
$$
$$
[F^{(n)}_3,F^{(1)}_{\pm}]=\pm F^{(n)}_{\pm}=[F^{(1)}_3,F^{(n)}_{\pm}] ,\;\;\;\;\;
[F^{(n)}_3,E^{(1)}_{\pm}]=0=[F^{(1)}_3,E^{(n)}_{\pm}]
$$
\begin{equation}
\label{eq33}
[E^{(n)}_+,E^{(1)}_-]=E^{(n)}_3=[E^{(1)}_+,E^{(n)}_-] ,\;\;
[F^{(n)}_+,F^{(1)}_-]=F^{(n)}_3=[F^{(1)}_+,E^{(n)}_-]
\end{equation}

$$
[\tilde{E}^{(n)}_3,U^{(1)}_{\pm}]=\pm \tilde{U}^{(n)}_{\pm}
=[E^{(1)}_3,\tilde{U}^{(n)}_{\pm}] ,\;\;
[\tilde{E}^{(n)}_3,V^{(1)}_{\pm}]=\pm \tilde{V}^{(n)}_{\pm}
=[E^{(1)}_3,\tilde{V}^{(n)}_{\pm}]
$$
$$
[\tilde{E}^{(n)}_3,E^{(1)}_{\pm}]=\pm \tilde{E}^{(n)}_{\pm}
=[E^{(1)}_{3},\tilde{E}^{(n)}_{\pm}] ,\;\;\;\;\;
[\tilde{E}^{(n)}_3,F^{(1)}_{\alpha}]=0
=[E^{(1)}_{3},\tilde{F}^{(n)}_{\alpha}]
$$
$$
[\tilde{E}^{(n)}_{\pm},U^{(1)}_{\pm}]=0
=[E^{(1)}_{\pm},\tilde{U}^{(n)}_{\pm}] ,\;\;\;\;\;
[\tilde{E}^{(n)}_{\pm},U^{(1)}_{\mp}]=\mp \tilde{F}^{(n)}_{\pm}
=[E^{(1)}_{\pm},\tilde{U}^{(n)}_{\mp}]
$$
$$
[\tilde{E}^{(n)}_{\pm},V^{(1)}_{\pm}]=0
=[E^{(1)}_{\pm},\tilde{V}^{(n)}_{\pm}] ,\;\;\;\;\;
[\tilde{E}^{(n)}_{\pm},V^{(1)}_{\mp}]=\mp \tilde{F}^{(n)}_{\mp}
=[E^{(1)}_{\pm},\tilde{V}^{(n)}_{\mp}]
$$
$$
[\tilde{E}^{(n)}_+,E^{(1)}_-]=\tilde{E}^{(n)}_3-I^{(n)}_0
=[E^{(1)}_+,\tilde{E}^{(n)}_-] ,\;\;
[\tilde{F}^{(n)}_+,F^{(1)}_-]=\tilde{F}^{(n)}_3-I^{(n)}_0
=[F^{(1)}_+,\tilde{F}^{(n)}_-]
$$
$$
[\tilde{E}^{(n)}_{\pm},F^{(1)}_{\pm}]=\mp \tilde{V}^{(n)}_{\pm}
=[\tilde{F}^{(n)}_{\pm},E^{(1)}_{\pm}] ,\;\;
[\tilde{E}^{(n)}_{\pm},F^{(1)}_{\mp}]=\pm \tilde{U}^{(n)}_{\pm}
=[E^{(1)}_{\pm},\tilde{F}^{(n)}_{\mp}]
$$
$$
[\tilde{U}^{(n)}_+,U^{(1)}_-]=\tilde{E}^{(n)}_3-\tilde{F}^{(n)}_3
=[U^{(1)}_+,\tilde{U}^{(n)}_-] ,\;\;
[\tilde{V}^{(n)}_+,V^{(1)}_-]=\tilde{E}^{(n)}_3+\tilde{F}^{(n)}_3
=[V^{(1)}_+,\tilde{V}^{(n)}_-]
$$
$$
[\tilde{U}^{(n)}_{\pm},V^{(1)}_{\mp}]=\mp Y^{(n)}_{\mp}
=[U^{(1)}_{\pm},\tilde{V}^{(n)}_{\mp}] ,\;\;
[\tilde{U}^{(n)}_{\pm},V^{(1)}_{\pm}]=\mp X^{(n)}_{\pm}
=[\tilde{V}^{(n)}_{\pm},U^{(1)}_{\pm}]
$$
$$
[\tilde{F}^{(n)}_3,U^{(1)}_{\pm}]=\mp \tilde{U}^{(n)}_{\pm}
=[F^{(1)}_3,\tilde{U}^{(n)}_{\pm}] ,\;\;
[\tilde{F}^{(n)}_3,V^{(1)}_{\pm}]=\mp \tilde{V}^{(n)}_{\pm}
=[\tilde{V}^{(n)}_{\pm},F^{(1)}_3]
$$
$$
[\tilde{F}^{(n)}_3,F^{(1)}_{\pm}]=\pm \tilde{F}^{(n)}_{\pm}
=[F^{(1)}_{3},\tilde{F}^{(n)}_{\pm}] ,\;\;\;\;\;
[\tilde{F}^{(n)}_3,E^{(1)}_{\alpha}]=0
=[F^{(1)}_3,\tilde{E}^{(n)}_{\alpha}]
$$
$$
[\tilde{F}^{(n)}_{\pm},U^{(1)}_{\mp}]=0
=[F^{(1)}_{3},\tilde{U}^{(n)}_{\mp}] ,\;\;\;\;\;
[\tilde{F}^{(n)}_{\pm},U^{(1)}_{\pm}]=\mp \tilde{E}^{(n)}_{\pm}
=[F^{(1)}_{\pm},\tilde{U}^{(n)}_{\pm}]
$$
$$
[\tilde{F}^{(n)}_{\pm},V^{(1)}_{\pm}]=0
=[F^{(1)}_{\pm},\tilde{V}^{(n)}_{\pm}] ,\;\;\;\;\;
[\tilde{F}^{(n)}_{\pm},V^{(1)}_{\mp}]=\pm \tilde{E}^{(n)}_{\mp}
=[\tilde{V}^{(n)}_{\mp},F^{(1)}_{\pm}]
$$
$$
[\tilde{E}^{(n)}_3,E^{(1)}_3]=0
=[F^{(1)}_3,\tilde{F}^{(n)}_3] ,\;\;
[\tilde{E}^{(n)}_{\pm},E^{(1)}_{\pm}]=\mp X^{(n)}_{\pm},\;\;
[\tilde{F}^{(n)}_{\pm},F^{(1)}_{\pm}]=\mp Y^{(n)}_{\pm}
$$
$$
[\tilde{U}^{(n)}_{\pm},U^{(1)}_{\pm}]=0,\;\;
[\tilde{V}^{(n)}_{\pm},V^{(1)}_{\pm}]=0,\;\;
[X^{(n)}_{\pm},E^{(1)}_{\mp}]=\mp 2\tilde{E}^{(n)}_{\pm},\;\;
[X^{(n)}_{\pm},E^{(1)}_{\pm}]=0
$$
$$
[X^{(n)}_{\pm},U^{(1)}_{\mp}]=\mp 2\tilde{V}^{(n)}_{\pm},\;\;
[X^{(n)}_{\pm},U^{(1)}_{\pm}]=0,\;\;
[X^{(n)}_{\pm},V^{(1)}_{\mp}]=\mp 2\tilde{U}^{(n)}_{\pm},\;\;
[X^{(n)}_{\pm},V^{(1)}_{\pm}]=0
$$
$$
[X^{(n)}_{\pm},E^{(1)}_3]=\mp 2X^{(n)}_{\pm},\;\;
[X^{(n)}_{\pm},F^{(1)}_3]=0,\;\;
[X^{(n)}_{\pm},F^{(1)}_{\pm}]=0,\;\;
[X^{(n)}_{\pm},F^{(1)}_{\mp}]=0
$$
$$
[Y^{(n)}_{\pm},E^{(1)}_3]=0,\;\;
[Y^{(n)}_{\pm},F^{(1)}_3]=\mp 2Y^{(n)}_{\pm},\;\;
[Y^{(n)}_{\pm},E^{(1)}_{\pm}]=0,\;\;
[Y^{(n)}_{\pm},E^{(1)}_{\mp}]=0
$$
$$
[Y^{(n)}_{\pm},U^{(1)}_{\pm}]=\mp 2\tilde{V}^{(n)}_{\pm},\;\;
[Y^{(n)}_{\pm},U^{(1)}_{\mp}]=0,\;\;
[Y^{(n)}_{\pm},F^{(1)}_{\pm}]=0,\;\;
[Y^{(n)}_{\pm},F^{(1)}_{\mp}]=\mp 2\tilde{F}^{(n)}_{\pm}
$$
$$
[Y^{(n)}_{\pm},V^{(1)}_{\pm}]=0,\;\;
[Y^{(n)}_{\pm},V^{(1)}_{\mp}]=\pm 2\tilde{U}^{(n)}_{\mp},\;\;
[I^{(n)}_0,E^{(1)}_{\pm}]=\mp 2\tilde{E}^{(n)}_{\pm},\;\;
[I^{(n)}_0,F^{(1)}_{\pm}]=\mp 2\tilde{F}^{(n)}_{\pm}
$$
$$
[I^{(n)}_0,U^{(1)}_{\pm}]=0,\;\;
[I^{(n)}_0,V^{(1)}_{\pm}]=0,\;\;
[I^{(n)}_0,E^{(1)}_3]=0,\;\;
[I^{(n)}_0,F^{(1)}_3]=0
$$
$$
[\tilde{E}^{(n)}_{\alpha},I^{(1)}_0]=0,\;\;
[\tilde{F}^{(n)}_{\alpha},I^{(1)}_0]=0,\;\;
[\tilde{U}^{(n)}_{\pm},I^{(1)}_0]=0,\;\;
[\tilde{V}^{(n)}_{\pm},I^{(1)}_0]=0
$$
$$
[X^{(n)}_{\pm},I^{(1)}_0]=0,\;\;
[Y^{(n)}_{\pm},I^{(1)}_0]=0,\;\;
[I^{(n)}_0,I^{(1)}_0]=0,\;\;
[E^{(n)}_{\alpha},I^{(1)}_0]=0
$$
\begin{equation}
\label{eq34}
[F^{(n)}_{\alpha},I^{(1)}_0]=0,\;\;
[U^{(n)}_{\pm},I^{(1)}_0]=0,\;\;
[V^{(n)}_{\pm},I^{(1)}_0]=0 ,\;\; (\alpha=\pm,3)
\end{equation}
From Eqs.(34) we may know that $I^{(n)}_0$  is Casimor
operator.

\pagebreak

\end{document}